\documentstyle[twoside,fleqn,espcrc2,epsf,bm]{article}

\newcommand{\beq}{\begin{equation}}
\newcommand{\eeq}{\end{equation}}
\newcommand{\bea}{\begin{eqnarray}}
\newcommand{\eea}{\end{eqnarray}}

\newcommand{\order}[1]{{\cal O}(#1)}  

\title{ 
\vspace{-3.2cm} 
\begin{flushright}
{\small OHSTPY-HEP-T-02-010}\\
\end{flushright}
\vspace*{2.0cm}
$B_s$ Mesons using Staggered Light Quarks
} 

\author{
Matthew Wingate,\address{Department of Physics, The Ohio State University,
	Columbus, Ohio 43210, USA}\thanks{Talk delivered at 
{\sl Lattice 2002}, Boston}
Junko Shigemitsu,{}$^{\rm a}$
G.\ Peter Lepage,\address{Newman Laboratory of Nuclear Studies,
Cornell University, Ithaca, NY 14853, USA}
Christine Davies,\address{Department of Physics \& Astronomy,
University of Glasgow, Glasgow, G12 8QQ, UK}
and Howard Trottier\address{Physics Department, Simon Fraser University,
Burnaby, B.C., V5A 1S6, Canada }
} 

\begin{document}

\begin{abstract}
Last year we proposed using staggered fermions as the light
quarks, combined with nonrelativistic heavy quarks,
in simulations of heavy-light mesons.  A first round
of tests which focuses on the $B_s$ meson
has been completed using quenched lattices, and results are presented here
for the kinetic $B_s$ mass, the $B_s^* - B_s$ splitting, and 
$f_{B_s}$.  The next project, already underway,
is to compute the $B$ and $B_s$ decay constants and spectra 
on the $n_f = 2+1$ and $3$ MILC lattices.  We report on progress
with one set of these configurations.
\end{abstract}

\maketitle

\section{INTRODUCTION}

Although a variety of lattice methods have been used to simulate the
heavy quark within a heavy-light meson, the light
quark description has not been studied as thoroughly; only 
Wilson-like fermions have been used in large scale simulations to date.
Recent simulations of light hadrons suggest, however, that improved
staggered fermions are superior to Wilson-like fermions, particularly
for small quark masses.  Improved staggered fermions have an exact
chiral symmetry at zero mass, no $\order{a}$ errors and excellent
scaling, and they require significantly less computational expense.
These benefits will facilitate the study of leptonic decay constants
and form factors in semileptonic decays as the physical light quark limit
is approached.

We have been studying the coupling of staggered light
quarks to nonrelativistic heavy quarks for simulation
of heavy-light mesons.  Since our initial study last year
\cite{Wingate:2001mp} we have resolved a question regarding
the improved staggered action on a very coarse $8^3\times 20$
lattice and have
completed a calculation of the $B_s$ mass, $B_s^*-B_s$ mass
splitting, and $f_{B_s}$ on a finer $12^3\times 32$ lattice.
We present those results in Section \ref{sec:quenched}.  
In Section \ref{sec:unquenched} we discuss progress with
this simulation method on the $2+1$ flavor MILC configurations.

Given the constraints of this write-up we refer the reader to
\cite{Wingate:2001mp} for a reminder of our method for
constructing operators out of NRQCD and naive fields and to
\cite{Lepage:2001ym,Wingate:2001mp} for a discussion
of the constrained curve fitting we employ in extracting energies
and matrix elements.  A forthcoming paper \cite{Wingate:2002tba}
will give a detailed description of the methods used to obtain
the results presented here.

\section{QUENCHED RESULTS}
\label{sec:quenched}

\begin{table*}[ht]
\setlength{\tabcolsep}{1.5pc}
\newlength{\digitwidth} \settowidth{\digitwidth}{\rm 0}
\catcode`?=\active \def?{\kern\digitwidth}
\caption{\label{tab:fits}Summary of fits to pseudoscalar
heavy-light correlators.}
\begin{tabular*}{\textwidth}
{@{}c@{\extracolsep{\fill}}l@{\extracolsep{\fill}}cccc} \hline
volume & action & $1/a_s$ (GeV) & $a_s/a_t$ & $\chi^2_{\rm aug}$/DoF
 & $E_{\rm sim}({\bf p}=0)$ (MeV) \\ \hline
~~$8^3\times 20 $ & 1-link & 0.8 & 1.0 & 0.59 & 735(10) \\ 
~~$8^3\times 20 $ & AsqTad & 0.8 & 1.0 & 8.93 & -- \\ 
~~$8^3\times 48 $ & AsqTad & 0.7 & 5.3 & 1.59 & 790(36) \\ 
~~$8^3\times 48 $ & AsqTad w/o ${\hat{t}}$-Naik & 0.7 & 5.3 
& 1.03 & 901(19) \\ 
$12^3\times 32$ & 1-link & 1.0 & 1.0 & 0.48 & 873(9) \\ 
$12^3\times 32$ & AsqTad & 1.0 & 1.0 & 0.96 & 765(9) \\ 
\hline
\end{tabular*}
\end{table*}

\begin{figure}[t]
\vspace{4.5cm}
\includegraphics{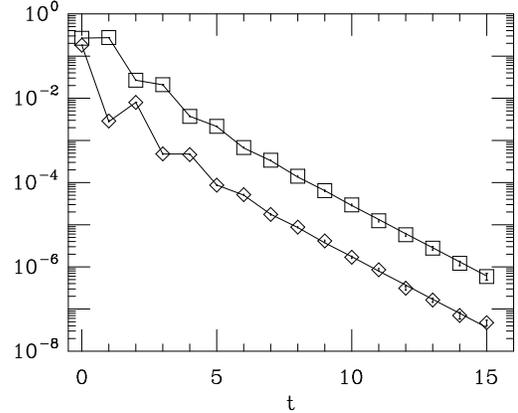}
\caption{
\label{fig:j0ratio} Squares show the $J_0^{(0)\dagger}(t)J_0^{(0)}(0)$
correlators on the $12^3\times 32$ lattice, and diamonds show
the $J_0^{(0)\dagger}(t)J_0^{(1)}(0)$ correlator times $-1$.}
\end{figure}

The quenched configurations were generated using the 
tadpole-improved tree-level Symanzik action.  Last year
we reported difficulty in obtaining reasonable fits 
for heavy-light correlators on a lattice with $1/a=0.8$ GeV 
when the $\order{a^2}$ staggered
action (AsqTad) was used \cite{Wingate:2001mp}.  By
the process of elimination, it was determined that the
extended 3-link hopping of the Naik term in the temporal direction
was responsible for the bad fits.  This term is known to give
rise to energetic negative-norm, ghost-like solutions to the
free fermion dispersion relation.  Since these states have energies
proportional to $1/a_t$, their effects on correlators should become
negligible as $a_t$ decreases.  Indeed on an
anisotropic lattice with $1/a_t = 3.7$ GeV
reasonable fits are obtained with
and without the temporal Naik term (see Table \ref{tab:fits}).  
Similarly, on an isotropic 
lattice with $1/a = 1.0$ GeV, there is no significant
contamination from negative norm states; a fit is shown in
Fig.\ \ref{fig:j0ratio}.

\begin{figure}[t]
\vspace{4.9cm}
\includegraphics{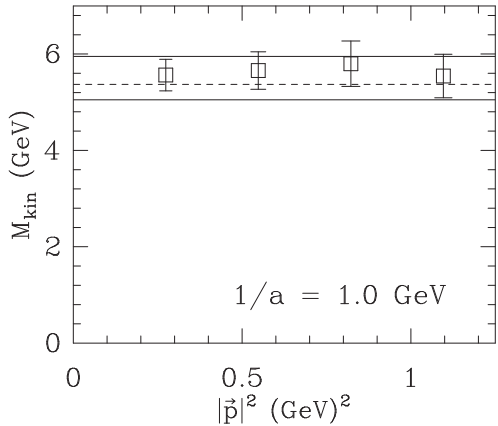}
\caption{
\label{fig:mkinps_q}
Kinetic $B_s$ mass on the quenched lattice.  The solid lines
mark the range of $M_{\rm pert}$ and the dashed line 
marks the experimental mass.}
\end{figure}

We focus now on results on the $12^3\times 32$ lattice, where
we used the AsqTad action with $am_0=0.10$ and an
$\order{\Lambda_{\rm QCD}/M_0, a^2}$-improved
NRQCD action with $aM_0=5.0$.  The bare quark masses were estimated to
coincide with the strange and bottom masses, respectively.
Meson masses can be computed using either finite momentum
correlators to construct the kinetic mass
\beq
M_{\rm kin} ~ =~  \frac{|\,{\bf p}|^2 - [E_{\rm sim}({\bf p})
-E_{\rm sim}(0)]^2}
{2[E_{\rm sim}({\bf p})-E_{\rm sim}(0)]} 
\eeq
or using perturbation theory
\beq
M_{\rm pert} ~=~ E_{\rm sim}(0) + Z_M M_0 - E_0 
\eeq
where $Z_M$ is the heavy quark mass renormalization and
$E_0$ is the self-energy constant.
For this lattice and $aM_0 = 5.0$ we find
$Z_M M_0 -E_0 = M_0 - 0.890\,\alpha_s + M_0\order{\alpha_s^2}$.
The result is
$M_{\rm pert} = 5.51 \pm 0.45$ GeV where the $\order{\alpha_s^2}$
uncertainty leads to the quoted error.
The comparison of $M_{\rm kin}$ to $M_{\rm pert}$ is shown in
Fig.\ \ref{fig:mkinps_q}.  

Mass splittings can be computed directly from differences of
simulation energies.  We find $M_{B_s^*} - M_{B_s} = 25 \pm 5$ MeV,
much smaller than the splitting expected from the experimental
$M_{B^*}- M_B = 46$ MeV, but in agreement with other quenched
results.

Matrix elements of the temporal QCD axial vector current $A_0$
are evaluated up through
$\order{\Lambda_{\rm QCD}/M_0, \alpha_s, \alpha_s/aM_0}$ 
by computing
matrix elements of 
\bea
J_0^{(0)} & = & \overline{q} \, \gamma_5\gamma_0 \, Q \\
J_0^{(1)} & = & -\frac{1}{2M_0}\overline{q}\, \gamma_5\gamma_0 \,
\bm{\gamma\cdot\nabla} \,Q 
\eea
and using the operator matching relation 
\beq
 A_0 ~\doteq~ Z_{A_0}J_0^{(0)}  +  J_0^{(1,{\rm sub})} 
\label{eq:axialvec}
\eeq
where the $\order{\Lambda_{\rm QCD}/M_0}$ contributions are absorbed 
into 
\beq
J_0^{(1,{\rm sub})} ~\equiv~ J_0^{(1)} - \alpha_s \zeta_{10}  J_0^{(0)} \, .
\eeq
With $aM_0=5.0$ 
we compute $Z_{A_0}=1+(0.208\pm0.003)\alpha_s$ and
$\zeta_{10} = -0.0997$.  Fits to appropriate local-local
correlators yield 
\beq
\frac{\langle 0 | J_0^{(1,{\rm sub})} | B_s\rangle}
{\langle 0 | J_0^{(0)} | B_s\rangle} ~=~ -0.034 \pm 0.004 ~{\rm (stat).}
\eeq
Applying (\ref{eq:axialvec}) and 
\beq
\langle 0 | A_0 | B_s\rangle ~\equiv~ f_{B_s}M_{B_s}
\eeq
gives the quenched result
\beq
f_{B_s} ~=~ 225 \pm 9 {\rm (stat)} \pm 20 {\rm (p.t.)}~{\rm MeV.} 
\eeq
The 20 MeV uncertainty is the estimate of the $\order{\alpha_s^2}$
error in $Z_{A_0}$.

\section{UNQUENCHED RESULTS}
\label{sec:unquenched}

Our work has just begun on the large suite of $2+1$ flavor
MILC ensembles (see e.g.\ \cite{Bernard:2001av}).  
So far we have simulated the $B_s$ system
on the configurations with $a\approx 0.13$ fm and dynamical
bare masses $a u_0 m_l = 0.01$ and $a u_0 m_s = 0.05$ (MILC
absorbs a factor of $u_0$ into their definition of $m_0$
compared to our convention). Our valence mass is
$au_0 m_0 = 0.05$.
We have used
the lattice scale determined from the $\Upsilon$ spectrum,
$1/a_\Upsilon = 1.58(2)$ GeV \cite{Davies:2002lat}. 
This scale is a bit higher than the one quoted in
\cite{Bernard:2001av}, $1/a_{r_1} = 1.46$ GeV;
however the heavy quark potential parameter $r_1$
is known only through quark models and so is not as rigorous
of a quantity for setting the lattice spacing.

\begin{figure}[t]
\vspace{4.9cm}
\includegraphics{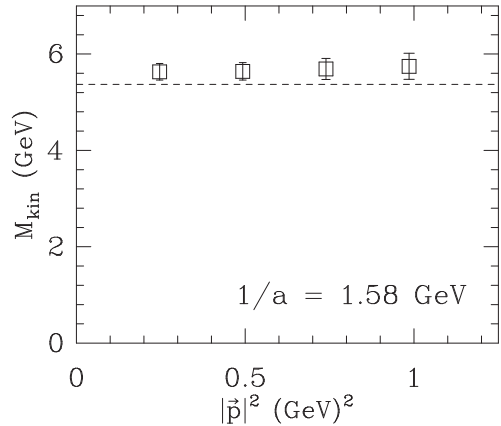}
\caption{
\label{fig:mkinps_unq}
Kinetic $B_s$ mass on the unquenched lattice.  The dashed line 
marks the experimental mass.}
\end{figure}

In Fig.\ \ref{fig:mkinps_unq} we show
the kinetic $B_s$ mass, e.g.\ the result for
${\bf p}=(0,0,1)$ gives $M_{\rm kin}(B_s) = 5.63(17)$ GeV.
We find that the unquenching dramatically improves the $B_s^* - B_s$
splitting, bringing it up to $42.5 \pm 3.7$ MeV.
The value we compute for $2M_{B_s}-M_\Upsilon = 1.361(8)(17)$ GeV
where the first error is dominated by the statistical error
in $E_{\rm sim}(B_s)$ and the second is due to the uncertainty 
in $a_\Upsilon$.
Using $a_\Upsilon$ implies that the strange sector actually corresponds
to $au_0m_0 \approx 0.04$ \cite{Hein:2002lat}.  From
the physical $M_{B_s}-M_B$
we estimate our value of $2M_{B_s}-M_\Upsilon$ obtained with 
$au_0 m_0 = 0.05$
is too high by about 45 MeV, 
or $2M_{B_s}-M_\Upsilon = 1.316(8)(17)$ GeV versus
1.278(5) GeV from experiment.  Simulations with smaller
$au_0m_0$ will permit a more rigorous interpolation to $M_{B_s}$.

\section*{ACKNOWLEDGMENTS}

We thank K.\ Foley and Q.\ Mason for their help
and the MILC collaboration for the use of their
gauge configurations.  
This work was supported by the DoE
and H.D.T.\ is supported by 
NSERC.



\begin{thebibliography}{15}

\bibitem{Wingate:2001mp}
M.~Wingate, J.~Shigemitsu and G.~P.~Lepage,
Nucl.\ Phys.\ Proc.\ Suppl.\  {\bf 106}, 379 (2002).

\bibitem{Lepage:2001ym}
G.~P.~Lepage {\it et al.}, 
Nucl.\ Phys.\ Proc.\ Suppl.\  {\bf 106}, 12 (2002)
hep-lat/0110175.

\bibitem{Wingate:2002tba}
M.~Wingate {\it et al.}, in preparation.



\bibitem{Bernard:2001av}
C.~W.~Bernard {\it et al.},
Phys.\ Rev.\ D {\bf 64}, 054506 (2001).

\bibitem{Davies:2002lat}
C.~T.~H.~Davies {\it et al.} and A.~Gray {\it et al.} in these proceedings.

\bibitem{Hein:2002lat}
J.~Hein {\it et al.} in these proceedings.

\end{thebibliography}
\end{document}